\documentclass[aps,twocolumn,groupedaddress,floatfix,superscriptaddress]{revtex4}
\usepackage{graphicx}
\usepackage{amsmath}
\usepackage{amstext}
\usepackage{amssymb}
\usepackage{amsfonts}
\usepackage{amsbsy}

\renewcommand{\v}[1]{\boldsymbol{#1}}

\newcommand{\Ref}[1]{Ref.~\cite{#1}}

\newcommand{\<}{\langle}
\renewcommand{\>}{\rangle}

\newcommand{\bpm}{\begin{pmatrix}}
\newcommand{\epm}{\end{pmatrix}}
\newcommand{\bmm}{\begin{matrix}}
\newcommand{\emm}{\end{matrix}}

\begin{document}
\bibliographystyle{apsrev}

\title{Lattice models for non-Fermi liquid metals}

\author{Michael Levin}
\affiliation{Department of Physics, University of California, Santa 
Barbara, California 93106}
\affiliation{Department of Physics, Harvard University, Cambridge,
Massachusetts 02138}

\author{T. Senthil}
\affiliation{Department of Physics, Massachusetts Institute of
Technology, Cambridge, Massachusetts 02139}


\date{\today}

\begin{abstract}
We present two $2D$ lattice models with non-Fermi liquid metallic
phases. We show that the low energy physics of these models is exactly
described by a Fermi sea of fractionalized quasiparticles coupled to a
fluctuating $U(1)$ gauge field. In the first model, the underlying degrees
of freedom are spin $1/2$ fermions. This model demonstrates that electrons
can in principle give rise to non-Fermi liquid metallic phases. In the
second model, the underlying degrees of freedom are spinless bosons. This
model provides a concrete example of a (non-Fermi liquid) Bose metal. With
little modification, it also gives an example of a critical $U(1)$
symmetric spin liquid.
\end{abstract}
\newcommand{\fig}[2]{\includegraphics[width=#1]{#2}}
\newcommand{\be}{\begin{equation}}
\newcommand{\ee}{\end{equation}}
\newcommand{\wdt}{\widetilde}
\newcommand{\red}{\textcolor{red}}
\newcommand{\ua}{\uparrow}
\newcommand{\da}{\downarrow}
\maketitle

\maketitle

\section{Introduction}

One of the basic tenets of conventional solid state physics is that clean
metals behave like Landau Fermi liquids at low energies. That is, they are
characterized by a sharp Fermi surface and sharp electron-like Landau
quasiparticles that are gapless at this surface. An extraordinarily
diverse collection of materials can be understood using this simple
framework.

However, a
number of experimental discoveries in the last two decades have challenged
our general description of metals as Landau Fermi liquids. The most
striking example is perhaps the normal state of the cuprate
superconductors. Breakdowns of Fermi liquid behavior have also been
observed in a number of other situations, notably in the vicinity of a
quantum critical point associated with the onset of magnetism in heavy
electron materials. \cite{CS0526,LRVW0715,GSS0886}

These and other experimental developments have strongly challenged the
conventional view of metals. Indeed a number of fundamental questions have
been brought into sharp focus. Do stable metallic phases of interacting
electrons exist that are not Landau Fermi liquids? Does a metal need to have
a sharp Fermi surface and/or sharp electron-like quasiparticles? A perhaps
even more fundamental question is whether metallicity is dependent on Fermi
statistics of the charge carriers. In other words, could a collection of
bosonic charge carriers form a metallic ground state? The possibility of
such a ``Bose metal" has been discussed much in the literature but without
adequate resolution. \cite{DD9951,DD0111,PD0343}

A limited amount of theoretical progress has been made towards answering
these questions. Much of this progress has come from slave particle
treatments of various models of correlated electron or boson systems. As is
well-known, this approach leads to a description of various phases of the
strongly correlated system in terms of fractionalized variables that are
coupled to emergent gauge fields. A number of non-Fermi liquid metallic
phases of interacting electrons have been described within such slave
particle gauge theories. \cite{IL8988, LN9221, L8980, KKLS0722,KKSS0828} It
is also possible to construct slave particle descriptions of Bose metal
phases of interacting bosons. \cite{MF0716} These constructions strongly
suggest that exotic non-Fermi liquid metallic phases of interacting
electrons or bosons are possible, at least in principle.

However, the slave particle approach has some well known
limitations. The most important of these is that it does not usually allow
for a definitive statement about the ground state of any particular
microscopic model. In practice this means that even though we may be able to
construct a stable effective field theory for a non-Fermi liquid phase using
slave particle variables, the approach does not allow us to identify
microscopic models that get into such a phase.

In this paper, we address this problem of finding definite microscopic
models that have non-Fermi liquid phases. We present two concrete lattice
models of non-Fermi liquid metals where the non-Fermi liquid physics can be
derived directly from the microscopic interactions. The first model,
a ``generalized Kondo lattice" model, is a two dimensional ($2D$) system 
of localized magnetic moments coupled to a separate band of conduction
electrons. The second model, a Bose-Hubbard model, is a $2D$ lattice boson
model with a ring exchange term.
While neither model is physically realistic, their low energy physics
is \emph{exactly} described by fractionalized fermionic
particles coupled to a fluctuating $U(1)$ gauge field. When the
fractionalized excitations form a Fermi sea the result is a non-Fermi
liquid metallic phase (modulo possible pairing instabilities at $T=0$).
\cite{IL8988, LN9221, L8980,HLR9312,P9417,AIM9448} The first model thus
demonstrates that electrons can form non-Fermi liquid metallic phases,
while the second model provides an example of a Bose metal - in fact a
non-Fermi liquid Bose metal. With little modification, it also gives an
example of a critical $U(1)$ symmetric spin liquid.

Our strategy for constructing and analyzing these models is very
straightforward. The idea is to make use of recent constructions
of exactly soluble lattice boson models with unusual
low energy physics. In particular, we make use of boson models
whose low energy excitations are fermionic quasiparticles coupled to
an emergent $U(1)$ gauge field. \cite{LWqed,LW0571} Previous work on these
models focused on the case where the fermionic excitations were gapped and the
system was insulating. Here, we simply modify the models so that the
fermionic excitations become gapless and open up a Fermi surface. In this
way, we construct a boson model with a (non-Fermi liquid) metallic phase. 
The construction of the fermion model is similar (and in some sense even 
easier since the fermionic quasiparticle excitations come for free).

We would like to mention that lattice models of
(non-Fermi liquid) Bose metals have appeared previously in the
literature. \cite{LWqed} The main contribution of this paper is that the
models presented here are simple and can be understood with minimal
background. In addition, we discuss their implications for non-Fermi liquid 
metallic phases, unlike previous work where the examples were simply 
mentioned in passing. 

The paper is organized as follows. In section II, we present
the ``generalized Kondo lattice model" while in section III, we describe
the boson model. In section IV, we summarize our results and conclude.

\section{Non-Fermi liquid metallic phase in a generalized Kondo lattice}
\begin{figure}[tb]
\centerline{
\includegraphics[width=2.0in]{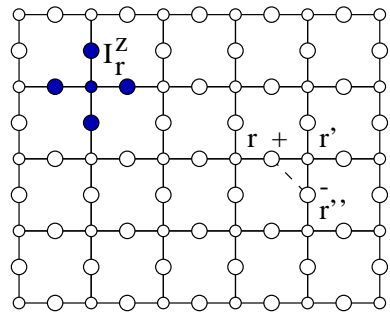}
}
\caption{
A picture of the cluster spin term $I^z_r = c^\dagger_r \sigma_z c_r +
\sum_{r' \in r} S^z_{rr'}$ and the spin-spin coupling term $S^+_{rr'}
S^{-}_{r'r''}$ in the generalized Kondo lattice model (\ref{kondoH}).
}
\label{kondo}
\end{figure}

The system we consider is a lattice of localized magnetic moments
coupled to a separate band of conduction electrons through a generalized
``Kondo lattice" Hamiltonian. The conduction electrons live on the sites of 
the square lattice while the local moments, which have spin $S$, live on the 
bonds. The Hamiltonian - a variant of Appendix B of \Ref{SVS0411} - is 
defined by
\begin{eqnarray}
H & = & H_t + H_{K\perp} + H_{Kz} + H_{s} \\
H_t & = & -t \sum_{\<rr'\>} \left(c^{\dagger}_{r\alpha} c_{r' \alpha} + h.c
\right) - \mu\sum_r c^\dagger_r c_{r} \nonumber \\
H_{K\perp} & = & J_K \sum_{\<rr'\>} \left[\left(c^{\dagger}_{r\ua}
c_{r'\da}  + c^{\dagger}_{r'\ua} c_{r\da} \right) S^{-}_{rr'} + h.c \right]
\nonumber \\
H_{Kz} & = & J_{Kz} \sum_r (I^z_r)^2 \nonumber \\
H_s & = & J_{\perp} \sum_{\<\<rr'r''\>\>} \left(S^+_{rr'} S^{-}_{r'r''} +
h.c \right) + J_{z} \sum_{\<rr'\>}\left( S^z_{rr'}\right)^2 \nonumber
\label{kondoH}
\end{eqnarray}
Here $c_{r\alpha}$ is the conduction electron destruction operator for site 
$r$ and spin $\alpha = \ua, \da$, while $\vec S_{rr'}$ is the local 
moment spin operator on the bond $\<rr'\>$. The operator $I^z_r$ in the 
$H_{Kz}$ term is defined by
\begin{equation}
I^z_r = c^\dagger_r \sigma_z c_r + \sum_{r' \in r} S^z_{rr'}
\end{equation}
where $r' \in r$ means that $r'$ is a nearest neighbor of $r$. $I^z_r$ may be
viewed as the total $z$-component of the spin associated with a cluster
composed of a site $r$ and the four bonds surrounding it (see Fig.
\ref{kondo}). Note that in defining this `cluster spin' we weight the
central electronic contribution twice as much as the contribution from each
of the four local moments on the bonds. This ensures that the total $z$
component of the spin is simply proportional to the sum of $I^z_r$ over all
clusters. Specifically
\be
\label{sztot}
S^z_{tot} = \frac{1}{2}\sum_r I^z_r
\ee
Finally, in the last term $H_s$, the notation $\<\<rr'r''\>\>$ refers to the
bonds shown in dashed lines in Figure \ref{kondo}.

These terms can be interpreted as follows. The first term $H_t$ contains 
the usual electron kinetic energy and chemical potential
terms. The second term $H_{K\perp}$ is a Kondo spin exchange between the 
$XY$ spin components of the local momenta and conduction electron systems. 
The structure of this term corresponds to an event where a conduction 
electron hops across a bond by flipping its spin along with an opposite spin 
flip of the local moment that resides on the bond. Clearly this term, as 
well as other the other terms described below, imply that the model only has 
symmetry under spin rotations about the $z$-axis of spin. The term $H_{Kz}$ 
with $J_{Kz} > 0$ penalizes fluctuations that change the total spin of the 
cluster that defines the operator $I^z_r$. Finally, the term $H_s$ is an 
inter-moment exchange term. Putting this all together, the Hamiltonian
(\ref{kondoH}) can be thought of as a generalized `Kondo-Heisenberg' model 
with a global $U(1)$ spin symmetry associated with conservation of the $z$
component of spin, and a distinct global $U(1)$ associated with
electric charge conservation.

We now argue that when $t, J_z, J_\perp \ll J_{Kz}$ and $J_z \ll J_K, 
\frac{J_\perp^2}{J_{Kz}}$ and $|\mu| < 4 J_K$, this
model realizes a non-Fermi liquid metallic phase (at least in the 
limit of large $S$). In order to simplify our analysis, we restrict to the 
case where the local moments have integer spin and we use a rotor description of 
this spin. Specifically we let
\begin{eqnarray}
S^{+}_{rr'} & \sim & e^{i\theta_{rr'}} \nonumber \\
S^z_{rr'} & = & n_{rr'}
\end{eqnarray}
where the phase $\theta_{rr'} \in [0, 2\pi]$ and `number' $n_{rr'}$ (which
can take any integer value) are canonically conjugate:
\be
[\theta_{rr'}, n_{rr'}] = i
\ee

We will focus exclusively on the limit of large $J_{Kz}$ as this allows us 
to access the non-Fermi liquid metallic phase. In that limit we first
diagonalize the $H_{Kz}$ term. The corresponding ground states satisfy
\be
\label{constraint}
I^z_r = c^\dagger_r \sigma_z c_r + \sum_{r' \in r} S^z_{rr'}~~~= 0
\ee
for each cluster $r$. This constraint leads to a highly degenerate manifold
of ground states. This degeneracy will be split by the other terms in the
Hamiltonian. Below we derive an effective Hamiltonian that lives entirely
within this degenerate ground state manifold. 

To that end we first make a
familiar change of notation to recast the constraint (\ref{constraint})
as the Gauss law constraint of a $U(1)$ gauge theory. We let $n_{rr'} =
\epsilon_r E_{rr'}, \theta_{rr'} = \epsilon_r a_{rr'}$ with $\epsilon_r
= \pm 1$ on the A and B sublattices of the square lattice. Further, we 
define new fermion operators through
\begin{equation}
f_r = \left\{ \bmm  c_r, & r \in A \\
\sigma^x c_r, & r \in B \emm \right.
\end{equation}
In this notation, the constraint (\ref{constraint}) becomes
\be
\label{gauss}
\vec \nabla \cdot \vec E + f^\dagger_r \sigma^z f_r = 0
\ee

We may now derive an effective Hamiltonian that lives entirely in this
constrained ground state subspace by degenerate perturbation theory. We
initially set $t = 0$ (and account for its effects later).
To leading order in $\left(J_{z}, J_K, \mu,
\frac{J_{\perp}^2}{J_{Kz}}\right)$, we find
\begin{eqnarray}
H_{\text{eff}} & = & H_f + H_{a} \\
H_f & = & J_K \sum_{\<rr'\>} \left[f^\dagger_{r}
e^{-ia_{rr'}\sigma^z}f_{r'} + h.c\right] \nonumber \\
& & - \mu\sum_r f^\dagger_r f_r \nonumber \\
H_a & = & -K \sum_P \cos\left(\vec \nabla \times \vec a\right) +
J_z \sum_{\<rr'\>} \vec E_{rr'}^2 \nonumber
\end{eqnarray}
where $K \sim \frac{J_{\perp}^2}{J_{Kz}}$ and the sum is taken over all
plaquettes $P$ of the square lattice. Including the $t$ term gives a 
four fermion interaction between the $f$-fermions at $O(t^2/J_{Kz})$. As we 
do not expect this small short distance interaction to be important, we will 
ignore it in what follows.

The effective Hamiltonian $H_{\text{eff}}$ describes two species of fermions
coupled to a fluctuating compact $U(1)$ gauge field with opposite gauge 
charges. The fermions carry physical charge $Q = e$, and spin $S^z = 0$. 
Depending on the value of $\mu$, the fermions can be gapped, or can open up
a Fermi surface.

Here, we are interested in the second case - which occurs for $|\mu| < 4
J_K$. In this case, $H_{\text{eff}}$ describes a Fermi sea of
two species of oppositely charged fermions coupled to a compact $U(1)$ 
gauge field. If we assume in addition that $J_z \ll K, J_K$, the instantons 
in the $U(1)$ gauge field will have a small fugacity and are expected to be 
irrelevant. \cite{IL8988,L0829} The $U(1)$ gauge field is then effectively 
non-compact at low energies.

This theory is precisely the low energy description of the holon metal phase
considered in \Ref{KKLS0722,KKSS0828} and in the $d$-wave correlated metals
of \Ref{MF0716}. The arguments given there as well as \Ref{IL8988, LN9221}
suggest that the resulting state is a metallic non-Fermi liquid with a
possible superconducting instability at low temperatures. A detailed
discussion of the physical properties of this state can be found in these
papers.

However, we would like to emphasize one particularly important
property here: the constructed metallic state has a \emph{spin gap}.
One way to see this is to note that the fermions carry spin $S^z = 0$
so that the state has no gapless spinful excitations. Alternatively, one
can see directly that the ground state manifold has $S^z_{tot} = 0$ due to
the constraint (\ref{constraint}) and the relation (\ref{sztot}).
Exciting states with $S^z_{tot} \neq 0$ costs energy of order $J_{Kz}$.
The nonzero spin gap implies that the electron tunneling
density of states will also have a gap of order $J_{Kz}$.

In summary, we have shown that the Kondo lattice model (\ref{kondoH})
realizes a non-Fermi liquid metallic phase in the regime where
$t, J_z, J_\perp \ll J_{Kz}$ and $J_z \ll J_K, \frac{J_\perp^2}{J_{Kz}}$ 
and $|\mu| < 4 J_K$. The low energy physics in this 
phase is described by a Fermi sea of fractionalized quasiparticles coupled 
to a fluctuating $U(1)$ gauge field. We would like to point out that this 
model also supports many other phases - including, for example, a Fermi 
liquid phase when $J_K, J_{Kz}, J_\perp \ll J_z \ll t$, and $|\mu| < 4 t$.

\section{A lattice boson model with a (non-Fermi liquid) metallic phase}

The model we consider is a lattice boson model where the bosons live on
the bonds of the square lattice. It is a 2D variant of the 3D quantum
rotor model discussed in \Ref{LWqed}. The Hamiltonian is given by

\begin{eqnarray}
H &=&
u \sum_{\< r s \>} n_{r s}^2 - \mu \sum_{\<r s\>} n_{rs} + U\sum_{r} N_{r}^2
\nonumber \\
& -  &  K\sum_{stuv} (\psi_{st}^{\dagger} \psi_{tu} \psi_{uv}^{\dagger}
\psi_{vs}
(-1)^{N_s}+h.c.)
\nonumber \\
&  - & t \sum_{\<\<rst\>\>} (\psi_{rs}^{\dagger}\psi_{st} (-1)^{N_{rst}} +
h.c.)
\label{twistedH}
\end{eqnarray}

\begin{figure}[tb]
\centerline{
\includegraphics[width=2.0in]{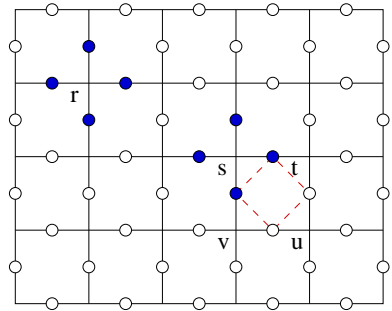}
}
\caption{
A picture of the cluster charge term $N_r= \sum_{r' \in r} n_{rr'}$ and
ring exchange term
$\psi_{st}^{\dagger} \psi_{tu} \psi_{uv}^{\dagger} \psi_{vs}(-1)^{N_s}$ in
(\ref{twistedH}).
}
\label{bosemodel}
\end{figure}

Here, $\psi_{rs}^{\dagger} = e^{i\theta_{r s}}$ denotes the boson creation
operator on the bond $\<rs\>$ and $n_{rs}$ is the corresponding boson
number operator. (As in the previous section, we will work in the
number-phase or quantum rotor representation of the bosons with
$[\theta_{rs}, n_{rs}] = i$).

The above Hamiltonian is essentially a boson ring exchange model with some
frustration. The $u$ term is an on-site repulsion term while $\mu$ is a
chemical potential. The $U$ term is a cluster charging term where the
clusters - labeled by $r$ - consist of the four bond-centered sites
$\<rr'\>$ neighboring a site $r$ of the square lattice (see Fig.
\ref{bosemodel}). The operator $N_r$ is the total boson number on these
four bonds:
\begin{equation}
N_r = \sum_{r' \in r} n_{rr'}
\end{equation}
The $K$ term is a ring exchange term involving four bond centered sites
adjoining a plaquette $stuv$. This is the usual ring exchange term
except for the additional phase
$(-1)^{N_s}$ which depends on the cluster charge on the upper left hand
corner of the plaquette, $s$ (see Fig. \ref{bosemodel}). This phase can be
thought of as some kind of frustration.

\begin{figure}[tb]
\centerline{
\includegraphics[width=1.5in]{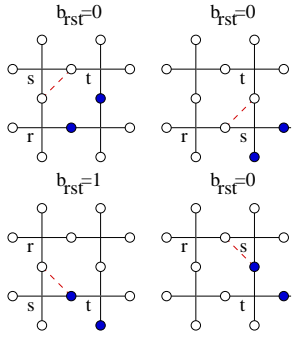}
}
\caption{
A picture of the boson hopping term $\psi_{rs}^{\dagger}\psi_{st}
(-1)^{N_{rst}}$. The operator $N_{rst} = b_{rst} + n_{rst}$ where $b_{rst}
= 0,1$ and $n_{rst}$ is the total boson number on two sites near $r,s,t$.
The value of $b_{rst}$ and the location of the two sites is different for
the four possible geometries of $r,s,t$. The sites are denoted above as
filled circles.
}
\label{bosehop}
\end{figure}

Finally, the $t$ term is a boson hopping term between neighboring bond
centered sites. Again, this is the usual boson hopping term except for the
additional phase $(-1)^{N_{rst}}$. Here, $N_{rst} = b_{rst} + n_{rst}$
where $b_{rst} =0,1$ depending on the specific geometry of $r,s,t$, and
$n_{rst}$ is the total boson number of two sites near $r,s,t$ - the
location of which also depends on the geometry of $r,s,t$ (see Fig.
\ref{bosehop}).

One way to understand the physics of this model is to consider a related
model without the additional factor of $(-1)^{N_s}$ in the ring exchange term
and $(-1)^{N_{rst}}$ in the hopping term. If we removed these factors, the
model would be very similar to the Bose-Hubbard models discussed in
\Ref{MS0204,Walight,MS0312,HFB0404}. Thus, like in
\Ref{MS0204,Walight,MS0312,HFB0404}
it could be mapped onto lattice $U(1)$ gauge theory coupled to bosonic 
charges. In brief, this mapping is given by setting $n_{rs} = \epsilon_r 
E_{rs}$, $\psi_{rs}^\dagger = b_r ^\dagger b_s ^\dagger e^{i \epsilon_r 
A_{rs}}$ where  $E_{rs}, A_{rs}$ are the lattice electric field and vector
potential, while $b_r^\dagger$ denotes the lattice boson creation operator.
Under this mapping the $u$ term maps onto an electric energy term
$E_{rs}^2$, the $K$ term maps onto a magnetic energy term $\cos(A_{st} +
A_{tu} + A_{uv} + A_{vs})$ and the $t$ term and $U$ terms map onto hopping
terms and mass terms for the bosons $b_{r}$.

The additional factors of $(-1)^{N_s},(-1)^{N_{rst}}$ modify this physics
in a simple way. First,
consider the factor of $(-1)^{N_s}$ in the ring exchange term. This factor
flips the sign of the ring exchange term if there is a boson $b_s$ at site 
$s$. Since this term corresponds to the magnetic energy term $\cos(A_{st} +
A_{tu}+A_{uv} + A_{vs})$ in the lattice gauge theory, this change in sign
means that the plaquette $stuv$ prefers to have a flux of $\pi$ instead of
$0$. Thus, the effect of the factor of $(-1)^{N_s}$ is to energetically
bind $\pi$ flux to the bosonic charges $b_s$.  The lowest energy charge 
excitation is therefore a composite of a bosonic charge and a $\pi$ flux. 
Similarly, one can see that
the effect of the $(-1)^{N_{rst}}$ in the $t$ term is to make the flux hop
together with the charges. Since a bound state of a bosonic charge and a
$\pi$ flux is a fermion, the net result is that the low 
energy effective theory for (\ref{twistedH}) maps onto a lattice model of 
\emph{fermions} coupled to a $U(1)$ gauge field instead of bosons.

In the following, we give a careful derivation of this result. We derive
the mapping to fermionic gauge theory from first principles. We then show
that the fermions can become gapless and open up a Fermi surface for
appropriate parameters - leading to a non-Fermi liquid metallic phase.
Specifically, we show that the non-Fermi liquid metallic phase occurs 
in the regime where $t \ll U$ and $u \ll K, t$ and $|U-\mu/2| < 8t$. 
 
We first suppose that $t = u = 0$ and then later consider the case of 
nonzero $t,u$. When $t = u = 0$, the Hamiltonian reduces to
\begin{displaymath}
H = U\sum_{\v r} (N_{\v r}-\mu/4U)^2 - K\sum_{\<stuv\>}
(\psi_{st}^{\dagger} \psi_{tu} \psi_{uv}^{\dagger} \psi_{vs} (-1)^{N_s}+h.c.)
\end{displaymath}
(where we have used the identity $\mu \sum_{\<rs\>} n_{rs} = \mu/2 \sum_r
N_r$).

This Hamiltonian is exactly soluble since the operators $\{N_r\},
\{\psi_{st}^{\dagger} \psi_{tu} \psi_{uv}^{\dagger} \psi_{vs}
(-1)^{N_s}\}$ all commute. Denoting the simultaneous eigenstate with $N_r
= n_r$, $\psi_{st}^{\dagger} \psi_{tu} \psi_{uv}^{\dagger} \psi_{vs}
(-1)^{N_s}= e^{i\phi_{stuv}}$ by $|n_r, \phi_{stuv}\>$, it is clear that
$|n_r, \phi_{stuv}\>$ is an eigenstate of the Hamiltonian with energy
\begin{equation}
E = U \sum_r (n_r-\mu/4U)^2 - 2K \sum_{\<stuv\>} \cos(\phi_{stuv})
\end{equation}
Assume that $\mu$ is close to, but
slightly less than $2U$. In that case the state $|n_r = 0, \phi_{stuv} =
0\>$ is the ground state. There are two types of low energy excitations:
``flux" excitations where $\phi_{stuv}$ is small but nonzero
for some plaquette $\<stuv\>$, and ``charge" excitations where $n_r =1$ for
some site $r$. The flux excitations are gapless while the charge
excitations have a small but finite gap $U - \mu/2$. Both types of
excitations are exact eigenstates and have no dynamics.

Now consider the case where $t$ is nonzero but much smaller than $U$. The
$t$ term will give dynamics to the charge excitations. Indeed,
one can see from the definition that when one applies the
operator $\psi_{rs}^{\dagger}\psi_{st} (-1)^{N_{rst}}$ to a state
$|\{n_r\},\{\phi_{stuv}\}\>$ it increases $n_r$ by $1$ and decreases $n_t$
by $1$. On the other hand, it does not change the fluxes $\phi_{stuv}$. We
conclude that this operator is an effective hopping term for charges: it
gives an amplitude for charges to hop from site $t$ to site $r$. Note that
this hopping term only moves charges within the $A$ and $B$ sublattices.
Thus, there are actually two species of charges - those on the $A$
sublattice and those on the $B$ sublattice.

While we know that the $t$ term gives an nonzero amplitude for charges to
hop, we need to understand the phases of these hopping matrix elements. In
particular, we need to understand the phases associated with (a) a charge
hopping around a plaquette, and (b) two charges exchanging places. We
begin with the problem of a charge hopping around a plaquette. We assume a
fixed flux configuration $\phi_{stuv}$ since the fluxes are completely
static.

\begin{figure}[tb]
\centerline{
\includegraphics[width=1.5in]{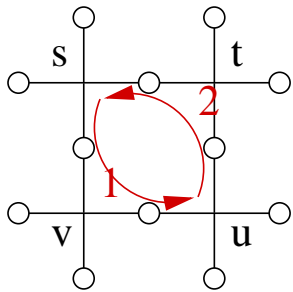}
}
\caption{
A charge at site $s$ has a nonzero matrix element to hop to site $u$ via
the operator $T_1 =\psi_{uv}^\dagger \psi_{vs} (-1)^{N_{uvs}}$. There is
also a nonzero matrix element to return to site $s$ via the operator $T_2
= \psi_{st}^\dagger\psi_{tu}(-1)^{N_{stu}}$. The total phase accumulated in
this process can be shown to be $e^{i\phi_{stuv}}$.
}
\label{plaqphase}
\end{figure}

Let us focus on a single plaquette $stuv$. We are interested in states
where the plaquette contains a single charge at one of the four sites,
$s,t,u,v$. We will label these states by $|s\>$, $|t\>$, $|u\>$, and
$|v\>$. Let us start with the state $|s\>$ where the
upper left hand corner $s$ is occupied. The term $T_1 =\psi_{uv}^\dagger
\psi_{vs} (-1)^{N_{uvs}}$ gives an amplitude for the charge to hop to the
lower right hand corner $u$. The term
$T_2=\psi_{st}^\dagger\psi_{tu}(-1)^{N_{stu}}$ gives an amplitude for the
charge to hop back to $s$ (see Fig. \ref{plaqphase}). The total phase
acquired by this process is given by
\begin{eqnarray}
e^{i\phi} &=& \<s|T_2 T_1 |s\> \nonumber \\
&=& \<s|\psi_{st}^\dagger \psi_{tu} (-1)^{N_{stu}}\psi_{uv}^\dagger
\psi_{vs} (-1)^{N_{uvs}}|s\>
\nonumber \\
&=& -\<s|\psi_{st}^\dagger \psi_{tu}\psi_{uv}^\dagger \psi_{vs}|s\>
\end{eqnarray}
where the last line follows from the fact that $N_{stu}+N_{uvs}= N_{u}+1
=1$ for this geometry.

We can rewrite this as:
\begin{eqnarray}
e^{i\phi} &=& -\<s|\psi_{st}^\dagger \psi_{tu}\psi_{uv}^\dagger
\psi_{vs}|s\> \nonumber \\
&=& \<s|\psi_{st}^\dagger \psi_{tu}\psi_{uv}^\dagger \psi_{vs}
(-1)^{N_s}|s\> \nonumber \\
&=& e^{i\phi_{stuv}}
\label{plaquettephase}
\end{eqnarray}

One can also show that the phase acquired when the charge starts in the
lower right hand corner $u$ is $e^{i\phi_{stuv}}$; when the charge starts
in the other two corners, one finds that the phase is $e^{-i\phi_{stuv}}$.
Thus, the charge excitations couple to the fluxes as if they carry a gauge
charge. Since the sign of the coupling is different when the charges are
at $s,u$ or $t,v$, we see that the charges on the $A$ and $B$ sublattices
carry \emph{opposite} gauge charges $\pm 1$.

\begin{figure}[tb]
\centerline{
\includegraphics[width=1.5in]{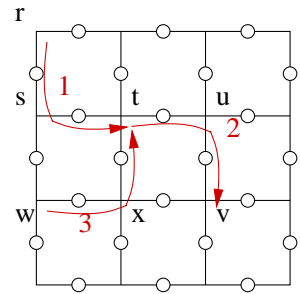}
}
\caption{
Two processes (A),(B) in which two charges starting at sites $r,w$ can
move to sites $t,v$. In process (A), the charge at $r$ hops to $t$ via
$T_1= \psi_{st}^\dagger\psi_{rs} (-1)^{N_{rst}}$ and from $t$ to $v$ via
$T_2=\psi_{uv}^\dagger\psi_{tu} (-1)^{N_{tuv}}$. The charge at $w$ then
hops to $t$ via $T_3= \psi_{tx}^\dagger\psi_{wx} (-1)^{N_{wxt}}$. In
process (B), the order of the hops is reversed: the charge at $w$ hops to
$t$ via $T_3$, and then to $v$ via $T_2$. The charge at $r$ then hops to
$t$ via $T_1$.
}
\label{hopalg}
\end{figure}

Next we compute the phase acquired when two charges exchange places. To
this end, we consider the two hopping processes (A),(B) shown in Fig.
\ref{hopalg}. The two processes both start with a state where there are
charges at sites $r$ and $w$, and end with a state with charges at sites
$t$ and $v$.

In process (A), the charge at site $r$ hops to $t$ via the hopping
operator $T_1 = \psi_{st}^\dagger\psi_{rs} (-1)^{N_{rst}}$, and then hops
from $t$ to $v$ via the hopping operator $T_2 =\psi_{uv}^\dagger\psi_{tu}
(-1)^{N_{tuv}}$. The charge at site $w$ then hops to $t$ via the operator
$T_3 = \psi_{tx}^\dagger\psi_{wx} (-1)^{N_{wxt}}$. In process (B), the
order of the three hops is reversed: the charge at site $w$ hopes to $t$
via the operator $T_3$ and then hops from $t$ to $v$ via the operator
$T_2$. Finally, the charge at site $r$ hops to $t$ via $T_1$.

Notice that the difference between the two processes is that in process
(A), the charges move from $r \rightarrow v$, and $w \rightarrow t$, while
in process (B), the charges move from $w \rightarrow v$, and $r
\rightarrow t$. Thus the difference between the two phases accumulated
in the two processes should tell us the phase associated with exchanging
the particles.

Simple algebra shows that
\begin{equation}
T_3 T_2 T_1 = -T_1 T_2 T_3
\label{hoppingalg}
\end{equation}
so that there is an extra phase of $\pi$ accumulated in process (A)
relative to process (B). We conclude that the charges are fermions.
(Indeed, the relation (\ref{hoppingalg}) is exactly the
``fermionic hopping algebra" from \Ref{LW0316}).

To summarize, we have shown that when $u = 0$, $t \ll U$, and $\mu$ is 
close to $2U$ the low energy physics of (\ref{twistedH}) is described by 
two species of fermionic quasiparticles carrying
opposite gauge charge minimally coupled to a compact $U(1)$ gauge field
at zero coupling constant (e.g. no $E^2$ term). In fact, the 
above calculations imply that the low energy physics of (\ref{twistedH}) 
can be \emph{exactly} mapped onto a
fermionic lattice gauge theory. (There is one technical issue with this
mapping, however. The corresponding lattice gauge theory
differs from standard lattice gauge theory in that it contains an
additional short distance interaction between fermions which occupy
neighboring sites. This interaction applies to the situation where a
fermion hops from site $t$ to site $r$ via $\psi_{rs}^\dagger\psi_{st}$
in the presence of another fermion at site $s$. As we do not expect this
short distance interaction to change the universal long distance physics,
we will ignore it here. In any case, if the reader is concerned about
this point, we would like to mention that this interaction can be
eliminated completely at the cost of using a more complicated model
Hamiltonian (\ref{twistedH}). See \Ref{LWqed} for a discussion).

Depending on the values of $\mu,t$ the fermions can be gapped or 
can open up a Fermi surface. Here, we are interested in the second case - 
which occurs for $|U -\mu/2| < 8 t$. In this case, the fermions will form
a (non-interacting) Fermi sea - occupying all single particle states up to 
energy $\mu/2 - U$.

Now imagine we turn on a small $u$, $u \ll K,t$. The $u$ term will
give dynamics to the flux configurations - formally $u$ corresponds to an
$E^2$ term for the gauge field. The result is thus a weak coupling
compact $U(1)$ gauge field coupled to a Fermi sea of two species of
oppositely charged fermions.

One can check that the fermions carry physical boson number (they each
carry boson number $1/2$). Thus, as in the previous example, we expect a
non-Fermi liquid metallic state with a possible superconducting pairing
instability at low temperature. The only difference is that the underlying
degrees of freedom here are bosons, not fermions. Thus, we have an example
of a non-Fermi liquid \emph{Bose} metal.

It is worth mentioning that this construction can easily be modified to
give an example of a critical $U(1)$ symmetric spin liquid. The first
step is to regard the boson model (\ref{twistedH}) as a quantum
rotor model, setting $L^z_{rs} = n_{rs}$, $L^+_{rs} =
\psi^{\dagger}_{rs}$, etc. One can then
obtain a spin $S$ spin model by replacing $L^z, L^+$ by $S^z, S^+$. When
$S$ is sufficiently large, we expect the spin model to be in the same
phase as the rotor model. This phase is a critical spin liquid described
by a Fermi sea of spinons coupled to a $U(1)$ gauge field.

\section{Conclusion}

In this paper, we have described two microscopic models of non-Fermi liquid
metals. In one model the underlying degrees of freedom are fermions; in
the other model, the basic degrees of freedom are bosons. The low energy
physics of both models is described by a Fermi sea of fractionalized
quasiparticles coupled to an emergent $U(1)$ gauge field.

We would like to emphasize that these microscopic models are far from unique. 
For example, similar models can be constructed (out of
either fermions or bosons) whose low energy physics is described by a
Fermi sea coupled to a $Z_2$ gauge field. These models are even
better controlled then the ones presented here, since the fermions are
completely non-interacting. There are many other possible constructions as 
well (such as modifications of Kitaev's exactly soluble honeycomb model
\cite{K062}). We hope that, collectively, these models can provide a
starting point for thinking about the microscopic physics of non-Fermi
liquid metallic phases.

\acknowledgments
This work was supported by NSF grants DMR-07-05255 and DMR-05-29399, and 
the Harvard Society of Fellows. 

\bibliography{nflmetal}

\newcommand{\noopsort}[1]{} \newcommand{\printfirst}[2]{#1}
  \newcommand{\singleletter}[1]{#1} \newcommand{\switchargs}[2]{#2#1}
\begin{thebibliography}{25}
\expandafter\ifx\csname natexlab\endcsname\relax\def\natexlab#1{#1}\fi
\expandafter\ifx\csname bibnamefont\endcsname\relax
  \def\bibnamefont#1{#1}\fi
\expandafter\ifx\csname bibfnamefont\endcsname\relax
  \def\bibfnamefont#1{#1}\fi
\expandafter\ifx\csname citenamefont\endcsname\relax
  \def\citenamefont#1{#1}\fi
\expandafter\ifx\csname url\endcsname\relax
  \def\url#1{\texttt{#1}}\fi
\expandafter\ifx\csname urlprefix\endcsname\relax\def\urlprefix{URL }\fi
\providecommand{\bibinfo}[2]{#2}
\providecommand{\eprint}[2][]{\url{#2}}

\bibitem[{\citenamefont{Coleman and Schofield}(2005)}]{CS0526}
\bibinfo{author}{\bibfnamefont{P.}~\bibnamefont{Coleman}} \bibnamefont{and}
  \bibinfo{author}{\bibfnamefont{A.~J.} \bibnamefont{Schofield}},
  \bibinfo{journal}{Nature} \textbf{\bibinfo{volume}{433}},
  \bibinfo{pages}{226} (\bibinfo{year}{2005}).

\bibitem[{\citenamefont{v.~Lohneysen et~al.}(2007)\citenamefont{v.~Lohneysen,
  Rosch, Vojta, and Wolfle}}]{LRVW0715}
\bibinfo{author}{\bibfnamefont{H.}~\bibnamefont{v.~Lohneysen}},
  \bibinfo{author}{\bibfnamefont{A.}~\bibnamefont{Rosch}},
  \bibinfo{author}{\bibfnamefont{M.}~\bibnamefont{Vojta}}, \bibnamefont{and}
  \bibinfo{author}{\bibfnamefont{P.}~\bibnamefont{Wolfle}},
  \bibinfo{journal}{Rev. Mod. Phys.} \textbf{\bibinfo{volume}{79}},
  \bibinfo{pages}{1015} (\bibinfo{year}{2007}).

\bibitem[{\citenamefont{Gegenwart et~al.}(2008)\citenamefont{Gegenwart, Si, and
  Steglich}}]{GSS0886}
\bibinfo{author}{\bibfnamefont{P.}~\bibnamefont{Gegenwart}},
  \bibinfo{author}{\bibfnamefont{W.}~\bibnamefont{Si}}, \bibnamefont{and}
  \bibinfo{author}{\bibfnamefont{F.}~\bibnamefont{Steglich}},
  \bibinfo{journal}{Nature Phys.} \textbf{\bibinfo{volume}{4}},
  \bibinfo{pages}{186} (\bibinfo{year}{2008}).

\bibitem[{\citenamefont{Das and Doniach}(1999)}]{DD9951}
\bibinfo{author}{\bibfnamefont{D.}~\bibnamefont{Das}} \bibnamefont{and}
  \bibinfo{author}{\bibfnamefont{S.}~\bibnamefont{Doniach}},
  \bibinfo{journal}{Phys. Rev. B} \textbf{\bibinfo{volume}{60}},
  \bibinfo{pages}{1261} (\bibinfo{year}{1999}).

\bibitem[{\citenamefont{Das and Doniach}(2001)}]{DD0111}
\bibinfo{author}{\bibfnamefont{D.}~\bibnamefont{Das}} \bibnamefont{and}
  \bibinfo{author}{\bibfnamefont{S.}~\bibnamefont{Doniach}},
  \bibinfo{journal}{Phys. Rev. B} \textbf{\bibinfo{volume}{64}},
  \bibinfo{pages}{134511} (\bibinfo{year}{2001}).

\bibitem[{\citenamefont{Phillips and Dalidovich}(2003)}]{PD0343}
\bibinfo{author}{\bibfnamefont{P.}~\bibnamefont{Phillips}} \bibnamefont{and}
  \bibinfo{author}{\bibfnamefont{D.}~\bibnamefont{Dalidovich}},
  \bibinfo{journal}{Science} \textbf{\bibinfo{volume}{302}},
  \bibinfo{pages}{243} (\bibinfo{year}{2003}).

\bibitem[{\citenamefont{Ioffe and Larkin}(1989)}]{IL8988}
\bibinfo{author}{\bibfnamefont{L.}~\bibnamefont{Ioffe}} \bibnamefont{and}
  \bibinfo{author}{\bibfnamefont{A.}~\bibnamefont{Larkin}},
  \bibinfo{journal}{Phys. Rev. B} \textbf{\bibinfo{volume}{39}},
  \bibinfo{pages}{8988} (\bibinfo{year}{1989}).

\bibitem[{\citenamefont{Lee and Nagaosa}(1992)}]{LN9221}
\bibinfo{author}{\bibfnamefont{P.~A.} \bibnamefont{Lee}} \bibnamefont{and}
  \bibinfo{author}{\bibfnamefont{N.}~\bibnamefont{Nagaosa}},
  \bibinfo{journal}{Phys. Rev. B} \textbf{\bibinfo{volume}{46}},
  \bibinfo{pages}{5621} (\bibinfo{year}{1992}).

\bibitem[{\citenamefont{Lee}(1989)}]{L8980}
\bibinfo{author}{\bibfnamefont{P.~A.} \bibnamefont{Lee}},
  \bibinfo{journal}{Phys. Rev. Lett.} \textbf{\bibinfo{volume}{63}},
  \bibinfo{pages}{680} (\bibinfo{year}{1989}).

\bibitem[{\citenamefont{Kaul et~al.}(2007)\citenamefont{Kaul, Kolezhuk, Levin,
  Sachdev, and Senthil}}]{KKLS0722}
\bibinfo{author}{\bibfnamefont{R.~K.} \bibnamefont{Kaul}},
  \bibinfo{author}{\bibfnamefont{A.}~\bibnamefont{Kolezhuk}},
  \bibinfo{author}{\bibfnamefont{M.}~\bibnamefont{Levin}},
  \bibinfo{author}{\bibfnamefont{S.}~\bibnamefont{Sachdev}}, \bibnamefont{and}
  \bibinfo{author}{\bibfnamefont{T.}~\bibnamefont{Senthil}},
  \bibinfo{journal}{Phys. Rev. B} \textbf{\bibinfo{volume}{75}},
  \bibinfo{pages}{235122} (\bibinfo{year}{2007}).

\bibitem[{\citenamefont{Kaul et~al.}(2008)\citenamefont{Kaul, Kim, Sachdev, and
  Senthil}}]{KKSS0828}
\bibinfo{author}{\bibfnamefont{R.~K.} \bibnamefont{Kaul}},
  \bibinfo{author}{\bibfnamefont{Y.~B.} \bibnamefont{Kim}},
  \bibinfo{author}{\bibfnamefont{S.}~\bibnamefont{Sachdev}}, \bibnamefont{and}
  \bibinfo{author}{\bibfnamefont{T.}~\bibnamefont{Senthil}},
  \bibinfo{journal}{Nature Physics} \textbf{\bibinfo{volume}{4}},
  \bibinfo{pages}{28} (\bibinfo{year}{2008}).

\bibitem[{\citenamefont{Motrunich and Fisher}(2007)}]{MF0716}
\bibinfo{author}{\bibfnamefont{O.~I.} \bibnamefont{Motrunich}}
  \bibnamefont{and} \bibinfo{author}{\bibfnamefont{M.~P.~A.}
  \bibnamefont{Fisher}}, \bibinfo{journal}{Phys. Rev. B}
  \textbf{\bibinfo{volume}{75}}, \bibinfo{pages}{235116}
  (\bibinfo{year}{2007}).

\bibitem[{\citenamefont{Halperin et~al.}(1993)\citenamefont{Halperin, Lee, and
  Read}}]{HLR9312}
\bibinfo{author}{\bibfnamefont{B.~I.} \bibnamefont{Halperin}},
  \bibinfo{author}{\bibfnamefont{P.~A.} \bibnamefont{Lee}}, \bibnamefont{and}
  \bibinfo{author}{\bibfnamefont{N.}~\bibnamefont{Read}},
  \bibinfo{journal}{Phys. Rev. B} \textbf{\bibinfo{volume}{47}},
  \bibinfo{pages}{7312} (\bibinfo{year}{1993}).

\bibitem[{\citenamefont{Polchinski}(1994)}]{P9417}
\bibinfo{author}{\bibfnamefont{J.}~\bibnamefont{Polchinski}},
  \bibinfo{journal}{Nucl. Phys. B} \textbf{\bibinfo{volume}{422}},
  \bibinfo{pages}{617} (\bibinfo{year}{1994}).

\bibitem[{\citenamefont{Altshuler et~al.}(1994)\citenamefont{Altshuler, Ioffe,
  and Millis}}]{AIM9448}
\bibinfo{author}{\bibfnamefont{B.~L.} \bibnamefont{Altshuler}},
  \bibinfo{author}{\bibfnamefont{L.~B.} \bibnamefont{Ioffe}}, \bibnamefont{and}
  \bibinfo{author}{\bibfnamefont{A.~J.} \bibnamefont{Millis}},
  \bibinfo{journal}{Phys. Rev. B} \textbf{\bibinfo{volume}{50}},
  \bibinfo{pages}{14048} (\bibinfo{year}{1994}).

\bibitem[{\citenamefont{Levin and Wen}(2006)}]{LWqed}
\bibinfo{author}{\bibfnamefont{M.}~\bibnamefont{Levin}} \bibnamefont{and}
  \bibinfo{author}{\bibfnamefont{X.-G.} \bibnamefont{Wen}},
  \bibinfo{journal}{Phys. Rev. B} \textbf{\bibinfo{volume}{73}},
  \bibinfo{pages}{035122} (\bibinfo{year}{2006}).

\bibitem[{\citenamefont{Levin and Wen}(2005)}]{LW0571}
\bibinfo{author}{\bibfnamefont{M.}~\bibnamefont{Levin}} \bibnamefont{and}
  \bibinfo{author}{\bibfnamefont{X.-G.} \bibnamefont{Wen}},
  \bibinfo{journal}{Rev. Mod. Phys.} \textbf{\bibinfo{volume}{77}},
  \bibinfo{pages}{871} (\bibinfo{year}{2005}).

\bibitem[{\citenamefont{Senthil et~al.}(2004)\citenamefont{Senthil, Vojta, and
  Sachdev}}]{SVS0411}
\bibinfo{author}{\bibfnamefont{T.}~\bibnamefont{Senthil}},
  \bibinfo{author}{\bibfnamefont{M.}~\bibnamefont{Vojta}}, \bibnamefont{and}
  \bibinfo{author}{\bibfnamefont{S.}~\bibnamefont{Sachdev}},
  \bibinfo{journal}{Phys. Rev. B} \textbf{\bibinfo{volume}{69}},
  \bibinfo{pages}{035111} (\bibinfo{year}{2004}).

\bibitem[{\citenamefont{Lee}(2008)}]{L0829}
\bibinfo{author}{\bibfnamefont{S.-S.} \bibnamefont{Lee}},
  \bibinfo{journal}{Phys. Rev. B} \textbf{\bibinfo{volume}{78}},
  \bibinfo{pages}{085129} (\bibinfo{year}{2008}).

\bibitem[{\citenamefont{Motrunich and Senthil}(2002)}]{MS0204}
\bibinfo{author}{\bibfnamefont{O.~I.} \bibnamefont{Motrunich}}
  \bibnamefont{and} \bibinfo{author}{\bibfnamefont{T.}~\bibnamefont{Senthil}},
  \bibinfo{journal}{Phys. Rev. Lett.} \textbf{\bibinfo{volume}{89}},
  \bibinfo{pages}{277004} (\bibinfo{year}{2002}).

\bibitem[{\citenamefont{Wen}(2003)}]{Walight}
\bibinfo{author}{\bibfnamefont{X.-G.} \bibnamefont{Wen}},
  \bibinfo{journal}{Phys. Rev. B} \textbf{\bibinfo{volume}{68}},
  \bibinfo{pages}{115413} (\bibinfo{year}{2003}).

\bibitem[{\citenamefont{Moessner and Sondhi}(2003)}]{MS0312}
\bibinfo{author}{\bibfnamefont{R.}~\bibnamefont{Moessner}} \bibnamefont{and}
  \bibinfo{author}{\bibfnamefont{S.~L.} \bibnamefont{Sondhi}},
  \bibinfo{journal}{Phys. Rev. B} \textbf{\bibinfo{volume}{68}},
  \bibinfo{pages}{184512} (\bibinfo{year}{2003}).

\bibitem[{\citenamefont{Hermele et~al.}(2004)\citenamefont{Hermele, Fisher, and
  Balents}}]{HFB0404}
\bibinfo{author}{\bibfnamefont{M.}~\bibnamefont{Hermele}},
  \bibinfo{author}{\bibfnamefont{M.~P.~A.} \bibnamefont{Fisher}},
  \bibnamefont{and} \bibinfo{author}{\bibfnamefont{L.}~\bibnamefont{Balents}},
  \bibinfo{journal}{Phys. Rev. B} \textbf{\bibinfo{volume}{69}},
  \bibinfo{pages}{064404} (\bibinfo{year}{2004}).

\bibitem[{\citenamefont{Levin and Wen}(2003)}]{LW0316}
\bibinfo{author}{\bibfnamefont{M.}~\bibnamefont{Levin}} \bibnamefont{and}
  \bibinfo{author}{\bibfnamefont{X.-G.} \bibnamefont{Wen}},
  \bibinfo{journal}{Phys. Rev. B} \textbf{\bibinfo{volume}{67}},
  \bibinfo{pages}{245316} (\bibinfo{year}{2003}).

\bibitem[{\citenamefont{Kitaev}(2006)}]{K062}
\bibinfo{author}{\bibfnamefont{A.}~\bibnamefont{Kitaev}},
  \bibinfo{journal}{Annals of Physics} \textbf{\bibinfo{volume}{321}},
  \bibinfo{pages}{2} (\bibinfo{year}{2006}).

\end{thebibliography}
\end{document}